\documentclass{ws-p8-50x6-00}
\begin{document}

%Definitions copied from newpasp.sty
\def\deg{\hbox{$^\circ$}}
\def\sun{\hbox{$\odot$}}
\def\earth{\hbox{$\oplus$}}
\def\la{\mathrel{\hbox{\rlap{\hbox{\lower4pt\hbox{$\sim$}}}\hbox{$<$}}}}
\def\ga{\mathrel{\hbox{\rlap{\hbox{\lower4pt\hbox{$\sim$}}}\hbox{$>$}}}}

\title{Origin and evolution of\\ neutron star magnetic fields}

\author{Andreas Reisenegger}

\address{Departamento de Astronom\'\i a y Astrof\'\i sica, Facultad de
F\'{\i}sica,\\
Pontificia Universidad Cat\'olica de Chile, Santiago, Chile
\\ areisene@astro.puc.cl}


\maketitle

\abstracts{This paper intends to give a broad overview of the
present knowledge about neutron star magnetic fields, their origin
and evolution. An up-to-date overview of the rich phenomenology
(encompassing ``classical'' and millisecond radio pulsars, X-ray
binaries, ``magnetars'', and ``thermal emitters'') suggests that
magnetic fields on neutron stars span at least the range
$10^{8-15}$ G, corresponding to a range of magnetic fluxes similar
to that found in white dwarfs and upper main sequence stars. The
limitations of the observational determinations of the field
strength and evidence for its evolution are discussed. Speculative
ideas about the possible main-sequence origin of the field
(``magnetic strip-tease'') are presented. Attention is also given
to physical processes potentially leading to magnetic field
evolution. }

\section{Introduction}

\indent

Given its role in opening the {\it International Workshop on
Strong Magnetic Fields and Neutron Stars,} this presentation has
the purpose of giving a general overview about what is currently
known about neutron star magnetic fields, their origin and
evolution. Inevitably, it will contain much of the same material
presented by the author in similar reviews in previous years
(e.g., Reisenegger 2001a), of which it should be regarded to be an
extension and update.

Magnetic fields are most likely the main form of ``hair'' that
allows neutron stars, contrary to black holes, to be distinguished
from each other and classified into phenomenologically very
different groups. Among single (or non-accreting binary) neutron
stars, we distinguish ``classical'' pulsars, millisecond pulsars,
soft gamma-ray repeaters, anomalous X-ray pulsars, and inactive,
thermal X-ray emitters (see, e.g., Becker \& Pavlov 2002). Binary
systems with mass transfer onto a neutron star can be divided into
high-mass and low-mass X-ray binaries (according to the companion
mass), with substantially different properties. Magnetic fields
play an essential role by accelerating particles, by channeling
these particles or accretion flows, by producing synchrotron
emission or resonant cyclotron scattering, and by providing the
main mechanism for angular momentum loss from non-accreting stars.
Moreover, evidence is mounting that soft gamma-ray repeaters and
anomalous X-ray pulsars are really only slightly distinct types of
very strongly magnetized neutron stars (``magnetars'') in which
the magnetic field is the main energy source for the observed
radiation.

On the other hand, we actually know surprisingly little about
neutron star magnetic fields. In particular, most ``measurements''
of neutron star magnetic fields are indirect inferences, which are
put in doubt both by their inconsistency with other observational
evidence and with plausible theoretical models for the physics of
their surroundings. Even less is known about the geometry of the
magnetic field, its evolution, and its origin, so there is open
space for speculation, modelling, and (hopefully) prediction of
measurable effects that might test the theoretical ideas.

Thus, I first review the observational ``classes'' of neutron
stars mentioned above, with a special eye on the evidence for the
presence and strength of the magnetic fields in each class (\S 2).
In \S 3, the inferred magnetic fluxes are compared to those of
other kinds of stars, and possible connections are discussed.
Section 4 surveys the evidence for and against magnetic field
evolution, and discusses physical processes which may lead to such
evolution. General conclusions are presented in \S 5.

\section{Classes of neutron stars and evidence for magnetic fields}

\subsection{Radio pulsars}

Radio pulsars are regularly pulsating sources of radio waves,
interpreted as magnetized, rotating neutron stars (Pacini 1967;
Gold 1968). Beams of radiation emerging from the poles of a
roughly dipolar magnetic field misaligned with respect to the
rotation axis appear as pulses every time they sweep the location
of the Earth. These pulses reveal rotation periods ($P$) from 1.55
milliseconds (ms) to several seconds. A very tangible illustration
of the impressively fast rotation rates (remember that more than a
solar mass is participating in this rotation!) is given in ``The
Sounds of Pulsars'', on the Jodrell Bank Observatory webpage
(http://www.jb.man.ac.uk/~pulsar/Education/Sounds/sounds.html).

Pulsar rotation periods are observed to lengthen with time ($\dot
P\equiv dP/dt>0$). The simplest model for the spin-down process is
to consider the neutron star as a magnetized body of moment of
inertia $I$, rotating in vacuum with angular velocity $\vec\Omega$
(Ostriker \& Gunn 1969). It loses rotational energy due to the
time-variation of its magnetic dipole moment vector $\vec\mu$,
which rotates at a fixed inclination $\alpha$ with respect to to
the rotation axis,

\begin{equation}
-{d\over dt}\left({1\over 2}I\Omega^2\right)={2\over 3
c^3}|\ddot{\vec\mu}|^2 ={1\over 6c^3}B^2R^6\Omega^4\sin^2\alpha,
\end{equation}

\noindent allowing us to infer the dipole magnetic field strength,
$B[{\rm G}]\approx 3.2\times 10^{19}\sqrt{P[{\rm s}]\dot P}$ (for
radius $R=10$ km, $I=10^{45}{\rm g\,cm}^2$, and
$\alpha=90^\circ$).

This ``dipole in vacuum'' model is unlikely to be very accurate,
as real pulsars are surrounded by a co-rotating, highly conducting
magnetosphere, and by an interstellar medium whose plasma
frequency is much higher than the expected radiation, which
therefore cannot propagate. Somewhat more realistic models (e.g.,
Gold\-reich \& Julian 1969) tend to roughly confirm the estimate
of $B$ (being less sensitive to $\alpha$), so this estimate is
generally used. Assuming a constant field strength and moment of
inertia, eq. (1) can be integrated backwards in time to give a
divergent rotation rate at a time
$\tau=\Omega/(-2\dot\Omega)=P/(2\dot P)$ before the present (at
which the spin parameters are to be evaluated), defining a
characteristic ``spin-down age'' for the pulsar.

In terms of these parameters, radio pulsars fall into two fairly
disjoint groups (e.g., Phinney \& Kulkarni 1994):

\begin{itemize}
\item young ($\tau\sim 10^{3-7}$ yr), relatively slow ($P\sim 16$ ms
to several seconds), and strongly magnetized ($B\sim 10^{11-13}$
G) ``classical'' pulsars, and

\item old ($10^{8-10}$ yr), fast (1.55 to several ms), and weakly
magnetized ($10^{8-9}$ G) ``millisecond'' pulsars.
\end{itemize}

Confirming that $\tau$ is related to true age, many of the
``youngest'' classical pulsars (and none of the millisecond
pulsars) are found to be associated with supernova remnants (which
disperse after $\sim 10^5$ yr). On the other hand, the vast
majority of pulsars in globular clusters are millisecond pulsars.
%(exceptions are described by Lyne et al. 1993, 1996).
An additional difference between the two classes is that most
millisecond pulsars are found in binary systems (in most cases
with old white dwarf companions), whereas the vast majority of
classical pulsars are single.

One problem with this general picture is that, in the few cases
where it is possible to measure $\ddot\Omega$ reliably (all of
which are young pulsars), the so-called ``braking index,''
$n\equiv\Omega\ddot\Omega/\dot\Omega^2$, does {\it not} agree with
the canonical $n=3$ predicted in the dipole spin-down model, but
has significantly {\it smaller} values, which differ from one
pulsar to another (see references in Kaspi et al. 2001). The
inclusion of higher multipoles (e.g., quadrupole electromagnetic
or gravitational radiation) worsens the problem. This means that
the {\it inferred} magnetic dipole moment in young pulsars {\it
increases} with time, possibly connecting young pulsars with
moderate inferred dipole moments to the slightly older
``magnetars'' with stronger inferred dipoles. Whether this
corresponds to a true increase of the star's dipole moment has not
been settled (Blandford 1994).

\subsection{X-ray binaries and the ``recycling'' paradigm}

Neutron stars with young, high-mass companions (``high-mass X-ray
binaries'' or HMXBs) tend to appear as X-ray pulsars, in which the
accreted material is presumably channelled by the magnetic field
onto the polar caps. In some cases, cyclotron features have been
found in the X-ray spectrum, corresponding to magnetic fields
$B\sim (1-4)\times 10^{12}$ G (e.g., Makishima et al. 1999; Coburn
et al. 2002, and references therein). Note that these spectral
features (found also in ``magnetars'' and ``thermal emitters'',
see below) are the only direct measurements of neutron star
magnetic fields, akin to the many measurements of magnetic fields
on white dwarfs and other stars. Assuming that these objects
differ from (similarly young) classical radio pulsars only by the
presence of the nearby companion, this would give evidence that
the field of neutron stars is organized on a relatively large
scale, so the surface field and the dipole field are of comparable
magnitude.

Older, low-mass companions tend to live with non-pulsating neutron
stars in ``low mass X-ray binaries'' (LMXBs), in which the field
is presumably not strong enough to channel the accretion flow. In
some cases, evidence for a fast rotation (and a non-zero magnetic
field) has been found in the form of quasi-periodic oscillations
and, more recently, true, highly coherent millisecond pulsations
(Wijnands \& van der Klis 1998; Galloway et al. 2002; for a review
of oscillations in LMXBs, see van der Klis 2000). These give
support to the long-standing paradigm that explains the puzzling
fast rotation of the evidently old MSPs in terms of ``recycling''
of these neutron stars in an LMXB (e.g., Bhattacharya \& van den
Heuvel 1991). The accretion of mass from a Keplerian disk
extending almost to the surface of the star would also transfer a
large amount of angular momentum, which could spin the star up to
the observed, fast rotation. It is speculated that the mass
transfer might also lead to the decay of the magnetic field to the
low value observed in MSPs. In two ``classical'' pulsars with
relatively fast rotation and relatively weak magnetic field,
``incomplete recycling'' might have occurred (Lyne et al. 1993,
1996).

\subsection{Magnetars}

Two intriguing kinds of astronomical objects have in recent years
found a likely interpretation as very highly magnetized neutron
stars (see Thompson 2000 for a review; Kouveliotou et al. 2003 for
a popular account):

\begin{itemize}
\item Soft gamma-ray repeaters (SGRs)
are objects which repeatedly emit bursts
of gamma-rays, in addition to persistent X-rays. For three of
these sources, regular pulses have been observed in the persistent
X-ray emission, allowing the measurement of a rotation period and
period derivative (Kouveliotou et al. 1998; Hurley et al. 1999).

\item Anomalous X-ray pulsars (AXPs) show persistent X-ray
emission, modulated at a stable, slowly lengthening period.
Contrary to the standard, {\it binary} X-ray pulsars (\S 2.3),
they show no evidence for a companion star (see Mereghetti \&
Stella 1995; van Paradijs, Taam, \& van den Heuvel 1995;
Mereghetti 2000).
\end{itemize}

Recently, bursts have been detected from two AXPs (Gavriil et al.
2002; Kaspi et al. 2003), making the connection even closer.
Differences remain in terms of X-ray spectra, burst frequency, and
timing stability, but it is not clear whether there is a dichotomy
or just a continuum of properties, with ``transition objects''
connecting those which are more characteristic of each class. All
measured periods lie in the narrow range $5-12$ s, and objects in
both classes have been claimed to be associated with supernova
remnants (see Gaensler et al. 2001 for references and a critical
discussion), arguing for an interpretation as young neutron stars.
Remnant ages are in rough agreement with characteristic ages
inferred from spin-down (but see J. Horvath's presentation in
these Proceedings).

The dipole fields inferred from the spin-down rate are
$10^{14-15}$ G (Kouveliotou et al. 1998; Hurley et al. 1999), much
larger than in previously known classical pulsars, though radio
pulsars with similar inferred dipole fields have recently been
found (Camilo et al. 2000; McLaughlin et al. 2003). In addition,
features in the X-ray spectra of both SGR 1806-20 (Ibrahim et al.
2002) and AXP 1RXS J170849-400910 (Rea et al. 2003), interpreted
as proton cyclotron resonance lines (Ibrahim et al. 2003),
indicate $B\sim 10^{15}$ G, in reasonable agreement with the
inferred dipole fields.

Perhaps most importantly, the persistent X-ray luminosity of these
objects is much larger than their inferred spin-down power.
Therefore, unlike the case of radio pulsars, rotation can {\it
not} be a significant energy source. It has long been suggested
that magnetic energy may be the ultimate source of both the bursts
and the persistent radiation (Duncan \& Thompson 1992; Paczynski
1992; Thompson \& Duncan 1995, 1996), but this would still require
a total magnetic energy significantly larger than inferred from
the dipole field, i.e., a buried and/or disordered magnetic flux.
In any case, the strong magnetic field may modify the radiation
transport in the surface layers, so that these objects radiate a
much larger fraction of their fossil heat in X-rays (as opposed to
neutrinos) than less magnetic neutron stars (van Riper 1988; Heyl
\& Hernquist 1997a, b). The presence of a light-element atmosphere
has a similar, even stronger effect (Chabrier et al. 1997; Heyl \&
Hernquist 1997a). Our own calculations (M. Riquelme et al., in
preparation) show that the Landau quantization of electron and
proton energy levels for plausible magnetic fields in these
objects are {\it not} strong enough to affect the neutrino
emissivity (see also Baiko \& Yakovlev 1999).

\subsection{Thermal emitters}

At the opposite end of the range of activity and, perhaps,
magnetic field strength are neutron stars detected exclusively
through their quiescent, thermal emission in X-rays and
(sometimes) optical radiation. Some of these have been found at
the centers of shell-type supernova remnants (SNRs; see Pavlov et
al. 2002 for a review), whereas other objects are isolated (see
Haberl 2003 and references therein).
%; see also Becker \& Pavlov 2002 and Becker \& Aschenbach 2002
%for a review covering all known kinds of pulsars and isolated neutron stars).
Objects in SNRs have
relatively high temperatures ($T\sim 0.2-0.7$ keV) and small
emitting areas (compared to the expected surface area of a neutron
star), whereas isolated objects are cooler ($T\sim 0.1$ keV) and
therefore fainter and only detectable in our immediate Galactic
neighborhood. That we nevertheless find several of these objects
indicates that they are very abundant, possibly representing the
``quiet majority'' of neutron stars in our Galaxy.

Beyond these general properties, this ``class'' of neutron stars
appears to be heterogeneous and remains a puzzle. In the source of
SNR RCW 103, a substantial, long-term flux variation has been
found, with a 6.4 h modulation interpreted in terms of accretion
from a faint binary companion (Sanwal et al. 2002a; see Pavlov et
al. 2002 for earlier references). Two other SNR sources show
periodic variations probably attributable to rotation, with
periods $P\sim 0.2-0.4$ s (Hailey \& Craig 1995; Zavlin et al.
2000), in the general ballpark of classical pulsars, but quite
slow for young pulsars in SNRs. In isolated sources, periods
$P\sim 8-23$ s have been found (Haberl et al. 1997; Haberl et al.
1999; Hambaryan et al. 2002; Haberl \& Zavlin 2002; Haberl et al.
2003), remarkably similar to those of ``magnetars'', whereas $\dot
P$ measurements for one source provide evidence for a somewhat
weaker magnetic field (Kaplan et al. 2002; Zane et al. 2002). Some
objects show strong spectral features (Sanwal et al. 2002b; Haberl
et al. 2003; Bignami et al. 2003), which might be indicating the
magnetic field strengths in these objects, although a unique
interpretation of the lines (e. g., proton vs. electron cyclotron
scattering) is still lacking. On the other hand, the flagship
object RX J1856.5-3754 (Walter et al. 1996) has not revealed any
evidence for periodicity or spectral features, in spite of intense
observational efforts (Burwitz et al. 2003 and references
therein). The same is true for some of its ``cousins'' (Motch et
al. 1999).

Therefore, the origin, evolution, and magnetic field of these
objects remain clouded in mystery. I will therefore refrain from
discussing them further, only noting that they may well provide
crucial input to our general understanding of neutron stars and
their magnetic fields in a not too distant future.

\section{Origin of the magnetic field}

\subsection{Stars as ${\cal R}-L$ circuits and ``flux freezing''}

Probably, all stars at all stages of their evolution have some
magnetic field, due to electronic currents circulating in their
interiors.

Naively, one might expect that such currents should decay over the
(microscopic) time scale $\tau_{\rm coll}$ in which an average
electron transfers its momentum to a more massive particle through
a Coulomb (or other) collision. However, any decrease in the
current $I$ implies a decrease of the magnetic flux $\Phi=cLI$
through the stellar equatorial plane, where $c$ is the speed of
light, $L\sim R/c$ is the star's self-inductance, and $R$ is its
radius. [Here and below, I use Gaussian cgs units; see, e. g.,
Jackson (1975).] According to Lenz's law, such a flux decline will
induce an emf $\varepsilon=-c^{-1}d\Phi/dt=-LdI/dt$ that tends to
keep the current going as prescribed by Ohm's law,
$\varepsilon={\cal R}I$. The resistance ${\cal R}$ can be
estimated in terms of a typical conductivity
$\sigma=n_ee^2\tau_{\rm coll}/m_e$ (where $n_e$, $-e$, and $m_e$
denote the electron concentration, charge, and mass) by ${\cal
R}\sim c/(\sigma R)$.

Thus, the star is well described by an electric circuit with an
inductance $L$ and a resistance ${\cal R}$ connected in series, in
which the current decays at such a rate that the induced emf is
always as strong as required to maintain the instantaneous current
against resistive decay. The exponential (``Ohmic'') decay time is
thus

\begin{equation}
\tau_{\rm Ohm}={L\over {\cal R}}\sim\sigma\left(R\over c\right)^2=
n_er_eR^2\tau_{coll},
\end{equation}

\noindent where $r_e=e^2/(m_ec^2)$ is the ``classical electron
radius''. Since the electron concentration is typically high, but
specially since stellar radii (even in the very compact neutron
stars) are large (e. g., compared to typical laboratory scales),
in general $\tau_{\rm Ohm}\gg\tau_{\rm coll}$ by many orders of
magnitude. Thus, stellar magnetic fields can persist for very long
times, being effectively ``frozen'' into the plasma.

Essentially the only way of changing the magnetic field
configuration is by ``deforming the circuit'', i. e., by
macroscopic displacements of the plasma, which can be thought of
as carrying the magnetic flux lines along. In particular, when a
star changes its radius, it could be expected to preserve its
enclosed magnetic flux, changing the magnetic field strength in
inverse proportion to its cross-sectional area, $B\propto R^{-2}$.

\subsection{Kinship}

Most, if not all, neutron stars descend from main sequence stars
with masses $M_{MS}\ga 8 M_\odot$, i.e., O and early B stars,
while lower-mass main sequence stars give rise to white dwarfs. A
fraction of early-type stars (Ap/Bp stars) have strong, highly
organized magnetic fields (see, e. g., the contributions of G.
Mathys, J.D. Landstreet, S. Bagnulo, and N. Piskunov in Mathys et
al. 2001). The same is true for a fraction of the white dwarfs,
which tend to be more massive than their non-magnetic counterparts
(Liebert et al. 2003), and therefore plausibly more closely
related to neutron stars.

It has long been known that the magnetic fluxes of magnetic white
dwarfs and neutron stars are similar (e.g., Ruderman 1972),
suggesting a common origin, possibly through flux conservation
during the evolution from some progenitor phase. Although much
more strongly magnetic objects have been discovered in recent
years, the most strongly magnetic main sequence stars (Ap/Bp stars
with $R\sim$ few $R_\odot$ and $B\sim 3\times 10^4$ G; e.g.
Landstreet 1992), white dwarfs ($R\sim 10^{-2}R_\odot$, $B\sim
10^9$ G; e.g., Wickramasinghe \& Ferrario 2000), and neutron stars
(magnetars with $R\sim 10^{-5}R_\odot$ and $B\sim 10^{15}$ G; see
\S 2.2) still turn out to have remarkably similar magnetic fluxes,
$\Phi=\pi R^2B\sim 10^{5.5}R_\odot^2$ G, despite vast differences
in size, density, and magnetic field strength. Lower limits on
magnetic fluxes can unfortunately not be compared, as the magnetic
fields of most non-degenerate stars and white dwarfs are too weak
to be detected. Of course, we may also not yet know the most
magnetic stars, if they are scarce or manifest themselves
phenomenologically in a way we have not yet identified.

\subsection{``Magnetic strip-tease'' hypothesis}

Another interesting point is as follows.

Early-type main sequence stars have convective cores and radiative
envelopes. The mass of the convective core, $M_{conv}$, is a
strongly increasing function of the total mass of the star,
$M_{MS}$ (e.g., Kippenhahn \& Weigert 1994). The mass of the
eventual compact remnant, $M_{rem}$, is a much more weakly
increasing function of $M_{MS}$ (e.g., Weidemann 2000). As
Thompson \& Duncan (1993) already realized, the two curves cross
at $M_{MS}\approx 3-4 M_\odot$ ($\sim$ A0 stars), where
$M_{conv}=M_{rem}\approx 0.7 M_\odot$, a plausible dividing line
between magnetic (massive) and non-magnetic (low-mass) white
dwarfs (cf. Liebert et al. 2003).

A possible interpretation is that, during the main sequence phase,
the field is generated by a dynamo process in the convective core
of most early-type stars. (A coherent, equipartition-strength
field filling the convective core of a $4M_\odot$ main sequence
star produces approximately the maximum flux estimated above.) If
it remains confined to the same region during the later stages of
evolution, then low-mass white dwarfs have a magnetized region
buried in their interior and covered by an unmagnetized envelope
(formerly part of the radiative envelope on the main sequence),
whereas massive white dwarfs and neutron stars form exclusively
from magnetized material, and therefore have a strong surface
field.

Thompson \& Duncan (1993) also pointed out that the high
metallicity-dependence of the size of the convective core of main
sequence stars may account for the fact that the mass of a white
dwarf does {\it not} uniquely determine whether it is detectably
magnetic or not. Of course, some questions would still remain
open:

- Why do (only) a small fraction of upper main sequence stars have
detectable surface fields?

- Might dynamo-generated fields be transported outward through the
stellar envelope, at any time during its evolution from the
beginning of the main sequence to the white dwarf stage?

\section{Magnetic field evolution}

\subsection{Observational evidence}

Several arguments point toward the possibility of an evolving
magnetic field in neutron stars:

%\begin{itemize}
%\item
1) Generally speaking, young neutron stars appear to have strong
magnetic fields $\sim 10^{11-15}$ G (``classical'' radio pulsars,
``magnetars'', X-ray pulsars), whereas old neutron stars have weak
fields $\la 10^9$ G (ms pulsars, low-mass X-ray binaries). If
these two groups have an evolutionary connection, their dipole
moment must decay. Millisecond pulsars are believed to have been
spun up to their fast rotation by accretion from a binary
companion, a remnant of which is in most cases still present
(e.g., Phinney \& Kulkarni 1994). The reduction in the magnetic
dipole moment may be a direct or indirect consequence of the
accretion process, or just an effect of age.

%\item
2) Studies of the pulsar distribution on the $P-\dot P$ diagram
(analogous to ``normal'' stellar population synthesis studies on
the HR diagram) have led to the claim that the magnetic torque
decays on a time scale comparable to the life span of
``classical'' pulsars (Gunn \& Ostriker 1970). This claim was
strengthened by the simultaneous consideration of pulsar space
velocities and their spatial distribution perpendicular to the
plane of the Galaxy (e.g., Narayan \& Ostriker 1990), but was
later put in doubt by other authors (e.g., Bhattacharya et al.
1992), whose more careful analysis leads to opposite results.

%\item
3) If magnetar emission is powered by magnetic energy (Thompson \&
Duncan 1996), then the rms magnetic field $\langle \vec
B^2\rangle^{1/2}$ must decay.

%\item
4) A possible explanation for the ``anomalous'' braking indices
$n<3$ in young neutron stars is that their magnetic dipole moment
increases with time.
%\end{itemize}

In the remainder of this section, I discuss the physical
mechanisms that may lead to such an evolution of the magnetic
field.

\subsection{Physics of spontaneous field evolution}

The composition of neutron star matter is still highly uncertain
(e.g., Lattimer \& Prakash 2000), but it seems almost inevitable
that it will contain both neutral particles (plausibly neutrons)
and charged particles (protons, electrons, and possibly others).
All particles are highly degenerate. The relativistic energies of
the electrons reduce their cross-section for colliding against
protons, and most of the phase space for final states is blocked
by the Pauli principle, leading to a high conductivity and
consequently to an Ohmic decay time (as in eq. 2) longer than the
age of the Universe (Baym, Pethick, \& Pines 1969b). Therefore,
little diffusion of the magnetic field can occur.

Can the magnetic field move {\it with} the fluid matter inside the
neutron star, driven by magnetic stresses or buoyancy forces? Not
in an obvious way. The matter in an equilibrium neutron star is
fully catalyzed, i.e., weak interactions have had time to bring
each fluid element into chemical equilibrium, minimizing its free
energy by distributing baryon number optimally among different
``flavors'' of particles. This optimal distribution is
density-dependent, giving rise to a {\it mechanically stable
composition gradient} (Pethick 1992; Reisenegger \& Gold\-reich
1992), regardless of the uncertainties in the composition
(Reisenegger 2001b). Even in the simplest and most favorable
scenario, in which the matter is mostly neutrons, with a small
($\sim 1\%$) ``impurity'' of protons and electrons, magnetic
stresses of order the ``impurity'' contribution to the fluid
stresses are required to overcome the stabilizing forces,
demanding a magnetic field $\ga 10^{17}$ G. At lower field
strengths, the magnetic stresses can only build up a small
chemical imbalance, and evolve on a timescale determined by the
weak interactions that reduce this imbalance.
% (see the first entry in Table 2).

This leads to the question of whether the magnetic field could
move only with the charged particles, leaving the neutral
particles behind. This question was addressed in a simple model
(Gold\-reich \& Reisenegger 1992) in which protons and electrons
move under the effect of electromagnetic forces through a static
and uniform neutral background, scattering against each other and
against this background. It leads to the following evolution law
for the magnetic field,

\begin{equation}
{\partial\vec B\over\partial t}=\nabla\times\left(\vec v\times\vec
B\right) +\gamma\nabla\times\left(-{\vec j\over n_e e}\times\vec
B\right) -\nabla\times\left({c\over\sigma}\vec j\right),
\end{equation}

\noindent where $\vec v$ is a weighted average velocity of all
charged particles, $\vec j=c\nabla\times\vec B/(4\pi)$ is the
electric current (due to {\it relative} motions of the charged
particles), $n_e$ is the density of protons and electrons, $e$ is
the proton charge, $c$ is the speed of light, $\sigma$ is an {\it
isotropic conductivity}, limited by inter-particle collisions, and
$\gamma$ is a dimensionless factor ($|\gamma|<1$) whose magnitude
and sign depends on the relative coupling of protons and electrons
to the neutral background. Each term on the right-hand side has a
familiar (astro-)physical interpretation, in turn:

%\begin{itemize}
%\item
1) Advection of the magnetic flux by a {\it flow of charged
particles,} i.e., {\it ambipolar diffusion,} familiar from star
formation (e.g., Shu et al. 1987): The bulk flow arises from
magnetic stresses or buoyancy forces acting on the charged
particles, and is impeded by inter-particle collisions. It can be
decomposed into two modes, one curl-free and one divergence-free,
the first of which will be choked by the chemical potential
gradients it builds up in the charged particles, and can only be
effective if weak interactions can reduce these gradients. Since
the driving forces are $\propto B^2,$ this term is $\propto B^3,$
becoming much more effective at high field strengths.

%\item
2) Advection of the magnetic flux by the electric current, or {\it
Hall effect:} This is a ``passive'' or ``kinematic'' effect, not
``driven'' by any forces and which by itself does not change the
magnetic energy. However, it is nonlinear ($\propto B^2$) and
could possibly lead to small-scale structures in the magnetic
field (Goldreich \& Reisenegger 1992), particularly in the solid
crust, where ambipolar diffusion cannot occur (see Shalybkov \&
Urpin 1997; Rheinhardt \& Geppert 2002; Hollerbach \& R\"udiger
2002 for samples of recent work).

%\item
3) The familiar {\it resistive} or {\it Ohmic diffusion}: Linear
in $B$; it is quite ineffective for a large-scale field, but may
play a role in dissipating small-scale structures created by the
other (nonlinear) processes.
%\end{itemize}

For the flows disturbing chemical equilibrium (bulk flow and
curl-free ambipolar diffusion), the timescale is set by weak
interactions, which also produce the early cooling of neutron
stars (through neutrino emission), and which are strongly
temperature dependent. Therefore, if $B\la 10^{17}$ G, the only
way in which these processes can be effective before the star
cools down is to keep it hot by some other mechanism, such as
dissipation of magnetic energy (e.g., Thompson \& Duncan 1996).
However, even this is not guaranteed to work, since most of the
dissipated energy will be emitted in the form of neutrinos. If the
field is strong enough to create a substantial chemical imbalance
($B\ga 10^{16}$ G), the enhanced neutrino emission may even lead
to faster cooling (Reisenegger 1995).

None of these mechanisms appear to be interesting at field
strengths and time scales relevant to classical or millisecond
radio pulsars, unless the magnetic field is confined to a thin
layer in the outer crust of the star, where the conductivity is
reduced and a combination of Hall drift and Ohmic dissipation may
become effective. In magnetars, the high field strength makes both
the Hall drift and the ambipolar diffusion quite fast, and their
interaction may lead to interesting dynamics, particularly if the
rms interior field is somewhat higher than the inferred dipole
field, as required from energetic arguments. The Hall
reorganization of the field in the crust may also lead to ``Hall
fracturing'' and therefore to dissipation (Thompson \& Duncan
1996), without need of invoking the ``Hall cascade'' to small
scales.

This discussion did not consider the formation of superfluid and
superconducting states, which probably occurs early in the
evolution of a neutron star (Baym, Pethick, \& Pines 1969a),
concentrating vorticity and magnetic flux into quantized ropes. At
moderate to low temperatures, the neutron star fluid will be much
more complicated than in the description given above (Mendell
1998). The effect of these complications on magnetic field
evolution are not yet well-understood, though much has been
speculated. I will refrain from further discussion of these
issues.

\subsection{Induced field evolution}

External agents may also change the magnetic field of a neutron
star:

%\begin{itemize}
%\item
1) The strong thermal gradient in a cooling protoneutron star can
overcome the stratifying effect of the chemical gradient, leading
to convection. At the same time, the star has not had time to
transport angular momentum and will be differentially rotating.
This combination naturally acts as a dynamo, which is an
alternative to the ``fossil flux'' idea (discussed above) to give
rise to the magnetic field in neutron stars (Thompson \& Duncan
1993). It has the merit of having led to the {\it prediction} of
the existence of ``magnetars'' with field strengths $B\sim
10^{15}$ G.

%\item
2) The thermal gradient persists for a much longer time in the
outer crust of the star, where it may act as a battery, again
giving rise to a substantial field (Urpin \& Yakovlev 1980;
Blandford, Applegate, \& Hernquist 1983; Wiebicke \& Geppert 1996
and references therein). This may in principle explain an
increasing field in a young pulsar, as suggested by the braking
index measurements.

%\item
3) Accretion from a binary companion is a popular way of
decreasing the magnetic dipole moment, although there is no
agreement on the physics involved. Perhaps the most interesting
candidate process is the burial (``diamagnetic screening'') of the
magnetic flux by the accreted, highly conducting plasma
(Bisnovatyi-Kogan \& Komberg 1975; Romani 1993; Cumming et al.
2001). No full models of this process have been produced so far,
and three-dimensional simulations will eventually be needed to
make sure that all possible instabilities have been taken into
account. If effective, this process still begs the question of why
after its completion a minute, but fairly constant fraction of the
initial dipole moment is left or regenerated to be detectable in
ms pulsars. (The observed ms pulsars do {\it not} appear to be the
``tip of the iceberg'' of a distribution extending down to much
lower fields, since the death rate of their LMXB progenitors
already can barely account for the detectable ms pulsars; e.g.,
Phinney \& Kulkarni 1994; White \& Ghosh 1998.)
%\end{itemize}

\section{Conclusions}

Research on magnetic fields in neutron stars is undoubtedly in one
of its most interesting moments. Little is known about the
strength, structure, origin, and evolution of the field, but there
seems to be little doubt that it plays a fundamental role in
determining the increasingly rich phenomenology of these objects.
The coming years will most probably improve our understanding of
the ``magnetars'' and ``thermal emitters'', hopefully contributing
to a coherent picture of how these new subclasses fit together
with the more traditional radio pulsar and X-ray binary groups and
with other kinds of stars. Along the way, we may expect surprises
in many different wavelength bands and much exciting physics.

%\acknowledgments
\section*{Acknowledgments}

Peter Goldreich, Shri Kulkarni, and Marten van Kerkwijk are
thanked for extensive conversations and many insights, and Chris
Thompson for several comments and references that improved the
manuscript. The writing of this article was supported by FONDECYT
grant no. 1020840.

%\small

\end{document}